\documentclass{emulateapj}
\usepackage{graphicx}
\usepackage{natbib}

\def\simlt{\lower.5ex\hbox{$\; \buildrel < \over \sim \;$}}
\def\simgt{\lower.5ex\hbox{$\; \buildrel > \over \sim \;$}}
\def\solar{\ifmmode _{\mathord\ odot}\else $_{\mathord\odot}$\fi}
\def\msun{\ifmmode {\rm M}_{\mathord\odot}\else $M_{\mathord\odot}$\fi}
\def\lsun{\ifmmode {\rm L}_{\mathord\odot}\else $L_{\mathord\odot}$\fi}
\def\xpvec{x\lla*p{\lower1ex\hbox{$\scriptstyle\sim$}}_{\scriptscriptstyle\perp}}
\def\xvec{x\llap{\lower1ex\hbox{$\scriptstyle\sim$}}}
\def\qvec{q\llap{\lower1ex\hbox{$\scriptstyle\sim$}}}
\def\to{\ifmmode \rightarrow\else $\rightarrow$\fi}
\def\Mc{M\raise0.5ex\hbox{c}}

\def\none{\ifmmode ^{-1}\else $^{-1}$\fi}
\def\two{\ifmmode ^{2}\else $^{2}$\fi}
\def\ntwo{\ifmmode ^{-2}\else $^{-2}$\fi}
\def\three{\ifmmode ^{3}\else $^{3}$\fi}
\def\nthree{\ifmmode ^{-3}\else $^{-3}$\fi}
\def\four{\ifmmode ^{4}\else $^{4}$\fi}
\def\nfour{\ifmmode ^{-4}\else $^{-4}$\fi}
\def\five{\ifmmode ^{5}\else $^{5}$\fi}
\def\nfive{\ifmmode ^{-5}\else $^{-5}$\fi}

\def\g{\ifmmode {\rm g}\else g\fi}
\def\kg{\ifmmode {\rm kg}\else kg\fi}

\def\cm{\ifmmode {\rm cm}\else cm\fi}
\def\m{\ifmmode {\rm m}\else m\fi}
\def\km{\ifmmode {\rm km}\else km\fi}
\def\pc{\ifmmode {\rm pc}\else pc\fi}
\def\ly{\ifmmode {\rm ly}\else ly\fi}
\def\au{\ifmmode {\rm au}\else au\fi}

\def\s{\ifmmode {\rm s}\else s\fi}
\def\Hz{\ifmmode {\rm Hz}\else Hz\fi}
\def\y{\ifmmode {\rm y}\else y\fi}

\def\K{\ifmmode {\rm K}\else K\fi}
\def\ster{\ifmmode {\rm ster}\else ster\fi}
\def\erg{\ifmmode {\rm erg}\else erg\fi}
\def\dyn{\ifmmode {\rm dyn}\else dyn\fi}

\begin{document}

\title{The Shapes of Molecular Cloud Cores in Simulations and Observations}

\author{Stella S. R. Offner} \affil{Department of Physics, University of
California, Berkeley, CA 94720} 
\email{soffner@berkeley.edu}

\author{Mark R. Krumholz \altaffilmark{1}}
\affil{Department of Astrophysical Sciences, Princeton University, Princeton, NJ 08544
and
Department of Astronomy, University of California, Santa Cruz, CA 95064} 
\altaffiltext{1}{Hubble Fellow}

\begin{abstract}
In this study, we investigate the shapes of starless and protostellar cores using hydrodynamic, self-gravitating adaptive mesh refinement simulations of turbulent molecular clouds.
We simulate observations of these cores in dust emission, including realistic noise and telescope resolution,  and compare to the observed core shapes measured in Orion by \citet{nutter07}. 
The simulations and the observations have generally high statistical similarity, with particularly good agreement between simulations and Orion B. Although protostellar cores tend to have semi-major axis to semi-minor axis ratios closer to one, the distribution of axis ratios for starless and protostellar cores  are not significantly different for either the actual observations of Orion or the simulated observations. Because of the high level of agreement between the non-magnetic hydrodynamic simulations and observation, contrary to a number of previous
authors, one cannot infer the presence of magnetic fields from core shape distributions.
\end{abstract}

\section {Introduction}

A successful theory of star formation must explain certain basic characteristics of the early stages of cores and stellar natal conditions. One property of interest is the shape distribution of starless and protostellar cores, which is likely related to the the initial conditions of star formation such as the local magnetic field configuration, level of turbulence, and core collapse timescale 
(see review by \citealt{mckee07}). Strong magnetic fields in the early stages of core formation may either support gas perpendicular to the field lines yielding a distribution of oblate cores \citep{mous76} or compress the cores into a prolate geometry \citep{fiege00}.  The former argument assumes predominantly poloidal magnetic fields, whereas the latter work includes a toroidal field component resulting in a helical field geometry. 
Independent of magnetic fields, prolate cores may also arise as an artifact of filamentary cloud geometry, in which cores fragment at intervals according to the characteristic Jeans length of a cylinder \citep{hartmann02}.

Recent large scale turbulent simulations with and without magnetic fields tend to find distributions of cores that contain predominantly triaxial cores when viewed in 3D
\citep{klessen00, gammie03, basu04, li04, Offner08a}. In contrast, 
simplified numerical studies including ambipolar diffusion tend to find triaxial cores with an inclination towards oblateness \citep{basu04, ciolek06}. Observations of molecular cloud cores in various star forming regions tend to find projected core aspect ratios, $q=a/b$, around 2:1
\citep{myers91, jijina99, nutter07}, but comparison of these observations with simulations is hampered by the projection of the observed cores onto the plane of the sky. Some authors have presented analytic work attempting to overcome this difficulty by ``de-projecting" the observed cores statistically. For example, \citet{ryden96}, assuming axisymmetry, finds that cores are significantly more likely to be randomly oriented prolate objects than oblate objects. \citet[henceforth T07]{tassis07} utilizes a maximum-likelihood method to generate a distribution of ellipsoid axial ratios. He uses two base probability distribution functions and finds that oblate or triaxial cores agree well with observations of cores in Orion \citep[hereafter NWT]{nutter07}, results which are insensitive to the assumed underlying distribution. Overall, T07 finds that prolate cores are rare, and his method rules out a uniform distribution of oblate, prolate, and triaxial cores with greater than 99\% confidence.

Our study is complementary to this previous work, but we perform the core shape comparison in the observational domain rather than the theoretical one. This has a significant advantage over alternative approaches, because it allows us to realistically simulate the effects of finite telescope resolution and sensitivity, and to reduce the simulated observations and fit core shapes using the same methods used for the real data. In this paper, we post-process simulations in this manner and compare with the dataset of Orion reported by NWT, who collate and reanalyze various SCUBA observations of the Orion A and B North and South molecular cloud complexes. They report the masses and sizes for 393 cores. Consequently, this dataset not only concerns an interesting and well-studied region, but it is also sufficiently large for meaningful statistics. 

In section 2, we describe the details of the simulations and post-processing. We consider simulations where turbulence is continually driven and where it is allowed to decay.
In section 3,  we present comparisons between the simulations and observations and compare with T07.  In section 4, we summarize our conclusions. 

\section{Simulated Observations}

As described in \citet{Offner08a, Offner08b}, our two simulations are periodic boxes
containing an isothermal, non-magnetized gas that is initially not self-gravitating. After driving turbulent motions in the gas for two box crossing times, self-gravity is turned on.  In one simulation energy injection is halted and the turbulence gradually decays, while in the other turbulent driving is maintained so that the cloud
satisfies energy equipartition.
A sink particle is introduced when the Jeans conditions is exceeded on the finest AMR level, where the cell spacing is $\Delta x$=200 AU \citep{krumholz04}.  
Since isothermal self-gravitating gas is scale free, it is easy to normalize the simulations to the conditions observed in Orion using the thermal Jeans length and thermal Jeans mass (see scaling relations in \citealt{Offner08b}). For the simulation normalization, we adopt a gas temperature of $T=20$ K (NWT). We choose a gas density of $\rho = 9.74\times 10^{-21}$ g cm$^3$ (number density of hydrogen nuclei n$_{\rm H} = 4.2 \times 10^3$ cm$^{-3}$), corresponding to a simulation box length of 2 pc. 
At this density, the typical box column density is N$_{\rm H}=2.6 \times 10^{22}$ cm$^{-2}$, close to the measured central column density of the Orion B cloud, N$_{\rm H}=2.8 \times 10^{22}$ cm$^{-2}$ \citep{mad86, johnstone01}.  
Although \citet{mad86} observe a much larger region than the individual north and south complexes in Orion A and B, column densities are observed to be roughly scale-independent  \citep{larson81, heyer08a} (although it has been argued that this is a selection effect -- e.g.\ \citealt{ball02}). Thus, our normalization is consistent.
%

We pause at this point to add a caveat about our normalization: our box size is smaller than the individual Orion A and B North and South complexes, which have projected sizes of $\sim 3-4$ pc. Equivalently, our simulation contains a smaller number of total thermal Jeans masses of gas than the entire Orion A and B North and South complexes. The justification for this is that the periodic geometry of the simulation enables us to model a piece of the cloud rather than the entire cloud. As long as we have the correct mean density, the behavior of structures that are much smaller than the box size such as cores should not depend on the size of the simulation box. Indeed, the similarity between the results for driven and decaying turbulence, and the insensitivity of our results to our assumed density normalization (see below), seem to support the hypothesis that small-scale structure is insensitive to such large-scale features as the total box size and the number of thermal Jeans masses it contains. However, establishing this point definitively would require a suite of simulations with varying total sizes but the same resolution, and such a study is unfortunately too computationally costly to perform. We therefore simply caution readers on this point and proceed.

\begin{figure*}
\plotone{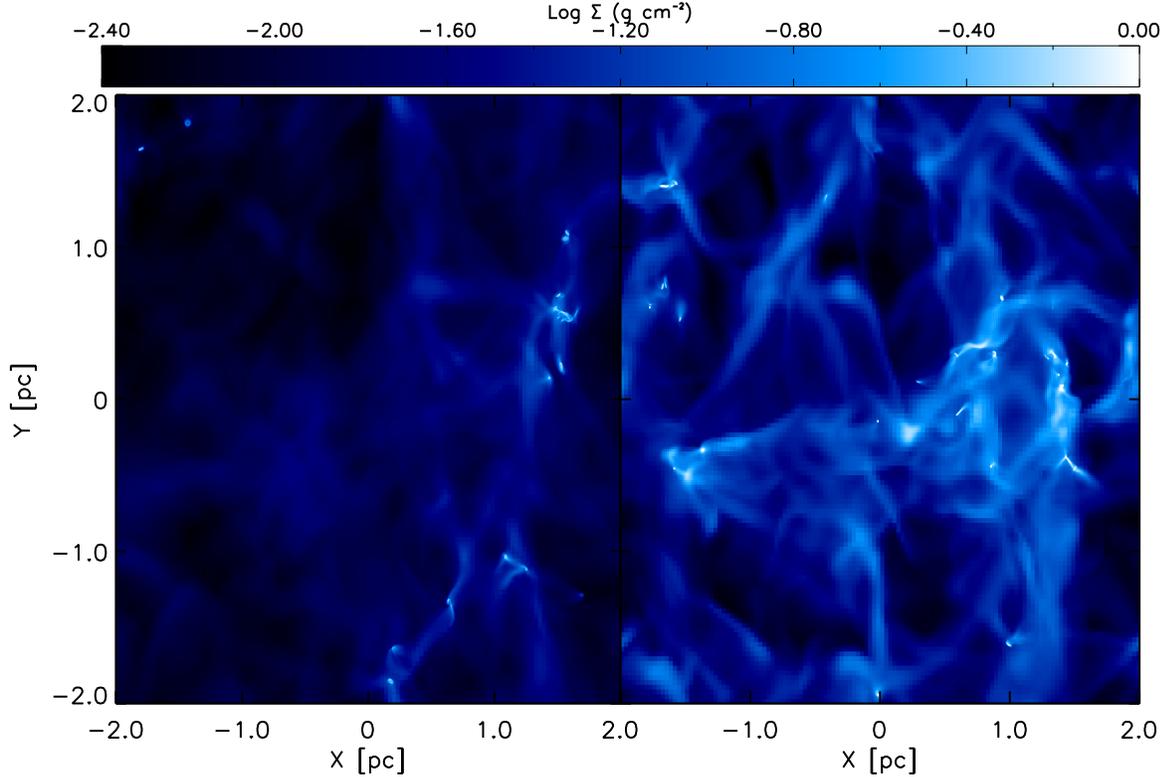} 
\caption{Logarithm of the column density, $\Sigma$, for one projection of the decaying (left) and driven(right) simulations at 1$t_{\rm ff}$.  \label{column}}
\end{figure*}

We run the simulations with gravity for a global freefall time, $t_{\rm ff}$,  and we compare with the observations at $t_{\rm ff}/2$ and $t_{\rm ff}$.
The column density of the two simulations at $t_{\rm ff}$ is displayed in Figure \ref{column}.
At these times, the Mach number, $\cal M$,  is 8.4 for the driven simulation and $\cal M=$ 5.3, 4.5 for the decaying simulation. 
For the driven simulation, this corresponds to half the Mach number of the larger Orion A and B regions that we compare with \citep{mad86}. However, it is unlikely that the high Mach number flow regions will have a significant effect on the details of the cores, which are generally subsonic to transonic.

It is possible to convert the simulation column density to an observed intensity using the relation
\begin{equation}
\label{NofI}
I_{\nu}= N_{\rm H} \Omega_{\rm mb} \mu_{\rm H} \kappa_{\nu} B_{\nu}(T_{\rm cloud})
\end{equation}
where $I_{\nu}$ is the flux density per beam at frequency $\nu$, $\Omega_{mb}$ is the solid angle subtended by the beam, $\mu_{H}$ is the mean mass per H atom, $\kappa_{\nu}$ is the dust opacity at frequency $\nu$, and B$_{\nu}$(T)
 is the Planck function \citep{enoch07}.  We set $\nu$=850 $\mu$m to match the observations of NWT, and following them we adopt $\kappa_{850}= 0.01$ cm$^2$g$^{-1}$. Note that eqn. \ref{NofI} assumes that the gas is optically thin at 850 $\mu$m. The densest sightline through the simulation has a column density of $\Sigma=0.6$ g cm$^{-2}$ or $\tau_{850}$=0.06, which is safely optically thin.

In order to facilitate comparison with NWT, we post-process the simulation data to have noise and resolution comparable to the SCUBA data set. In each projection direction, we integrate along the line of sight and convolve the column density image with a beam of  resolution 14''. We assume that the simulated cloud lies at a distance of 400 pc,  which is the average distance adopted by NWT.  
To each pixel in the smoothed image, we add a Gaussian noise distribution with $\sigma_{\rm NWT}$ = 20 mJy beam$^{-1}$, correlated over the FWHM size of the beam. This reproduces the coarser pixel resolution noise inherent in the SCUBA data.

Having generated a simulated column density map,  we analyze the data using the same procedure outlined in NWT, which we describe in the following steps.  
First, from the post-processed simulation data, we generate a large scale structure map by convolving the data with a beam size of 1'.  Second, we subtract this map from the high resolution version to remove large scale structure
to make core identification easier.
Third,  we define cores as density enhancements within $3\sigma_{\rm NWT}$ contours that contain a peak above 5$\sigma_{\rm NWT}$. With few exceptions, core shapes are generally elliptical. Finally, we fit an ellipse around each peak and match to the $3\sigma_{\rm NWT}$ contours ``by eye" guided by a sample eye-fit from NWT (D.\ Nutter, 2008, priv.\ comm.). 
An automated fitting algorithm would clearly be preferable to the ``by eye" procedure from the standpoint of reproducible comparison between different samples and authors. However, after experimenting with several options we were unable to find an algorithm that adequately reproduced the fits from NWT, while we were able to do so reasonably well by eye. Since the goal of this paper is to compare to the observed sample, we use by eye fits.
Based on the elliptical fits we assign each core an axis ratio $q$.  We identify cores that contain a sink particle within 0.05 pc of their core center as ``protostellar cores", while cores that do not contain a sink particle are ``starless".  In the sample of Orion cores, the protostellar cores are identified with the Spitzer IRAC camera.

\begin{figure}
\plotone{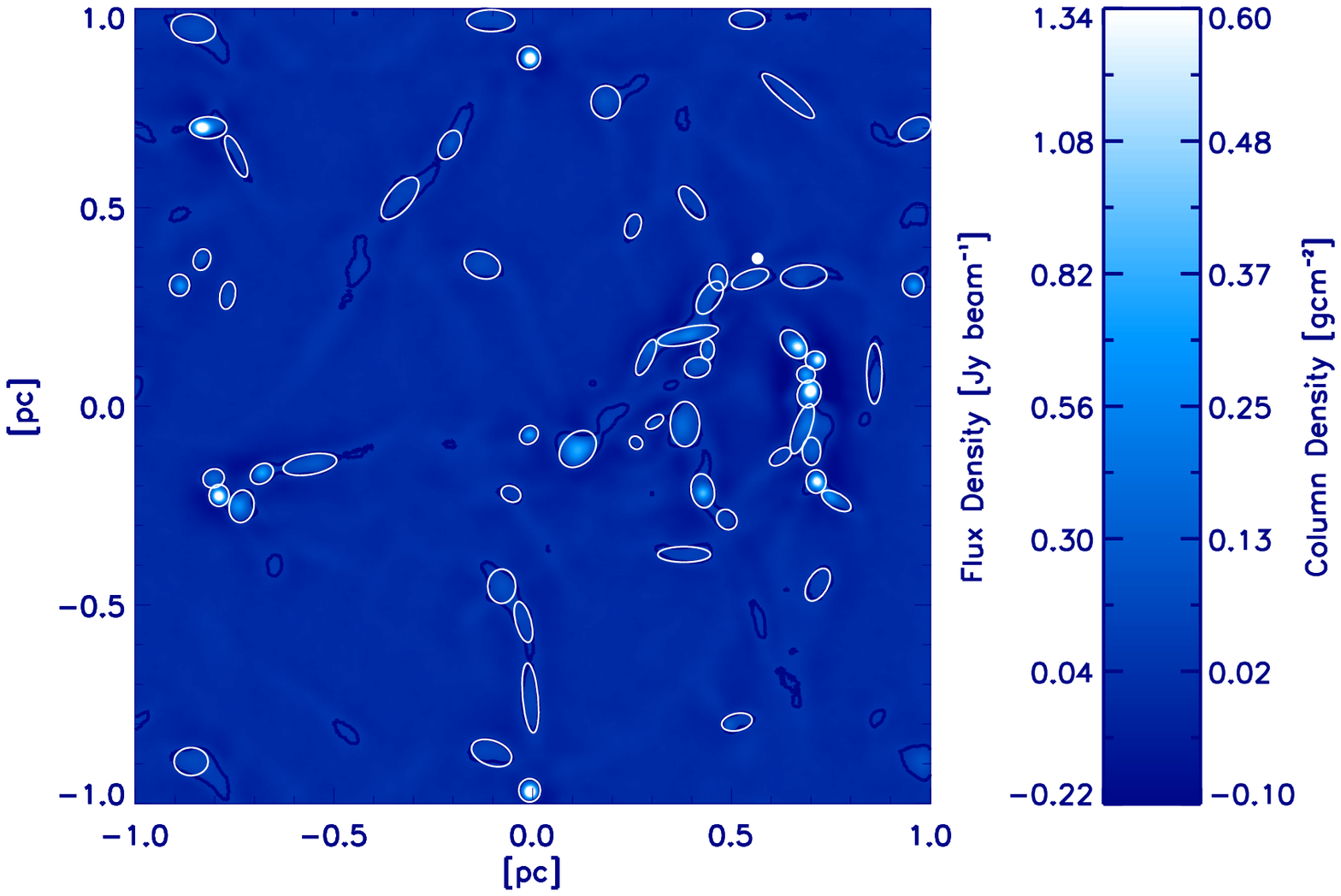} 
\caption{Column / flux density for one projection of the post-processed driven simulated cores at 1$t_{\rm ff}$. The image includes noise and beam-smearing. The $3\sigma_{\rm NT}$ contours are marked in black and the fitted ellipses are overlaid in white. \label{cores}}
\end{figure}

Figure \ref{cores} shows one projection through the simulation domain for the driven run, along with the bounding ellipses for all the cores we identify in that projection. 
For comparison, the unprocessed driven column density is shown in the right panel of Figure \ref{column}. 
To improve the statistics, we include cores from maps in each cardinal direction of the simulation. 
 The Kolmogov-Smirnov (KS) test provides a good statistical measure of the agreement of two distributions \citep{num92}. It derives 1 minus the confidence level at which the null hypothesis that the two were drawn from the same underlying distribution can be ruled out\footnotemark.
 \footnotetext{Formally, the two-sided K-S statistic we use is computed as follows. Consider two sets of $N$ and $M$ measurements of some quantity $x$ (in our case $x$ is the axis ratio, and the two sets of measurements are the simulated and observed values). Let $F_N(x)$ and $G_M(x)$ be the cumulative distribution functions for those measurements, i.e.\ $F_N(x)$ is the fraction of the $N$ measurements that yield a value less than or equal to $x$, and similarly for $G_M(x)$. The Kolmogorov-Smirnov statistic is then defined as $D_{N,M} = \mbox{sup}_{x} |F_N(x) - G_M(x)|$, i.e.\ the maximum distance between the two cumulative distribution functions. Kolmogorov's Theorem then states that we can reject the null hypothesis that $F_N(x)$ and $G_M(x)$ were drawn from the same parent distribution with a confidence level $\alpha$ if $\sqrt{NM/(N+M)} D_{N,M} > K_{\alpha}$, where $K_{\alpha}$ is defined implicitly by the equation $1-\alpha=\mbox{Pr}(K\leq K_{\alpha})$ and $\mbox{Pr}(K\leq x)\equiv 1 - 2\sum_{i}^{\infty} (-1)^{i-1}e^{-2i^2 x^2}$. Intuitively, $\alpha$ gives the probability that we could have measured a value of $D_{N,M}$ as large as we did if $F_N(x)$ and $G_M(x)$ were actually drawn from the same parent distribution. The smaller $\alpha$ is, the less likely that our samples would have produced such a large $D_{N,M}$ if the samples were drawn from the same parent distribution.}.
 In the remainder of the paper, we report the quantity 1 minus the confidence level multiplied by a factor of a 100 to give a percentage.
Using a KS test, we find that core samples from different lines of sight are consistent with being drawn from the same distribution (KS statistics $\sim 50\%$). This implies that the core ratios in the three projections are statistically indistinguishable, so our procedure of treating each orientation as an independent sample is consistent. The agreement among the orientations also shows that fitting by eye is reasonably reproducible.

\begin{figure}
\plotone{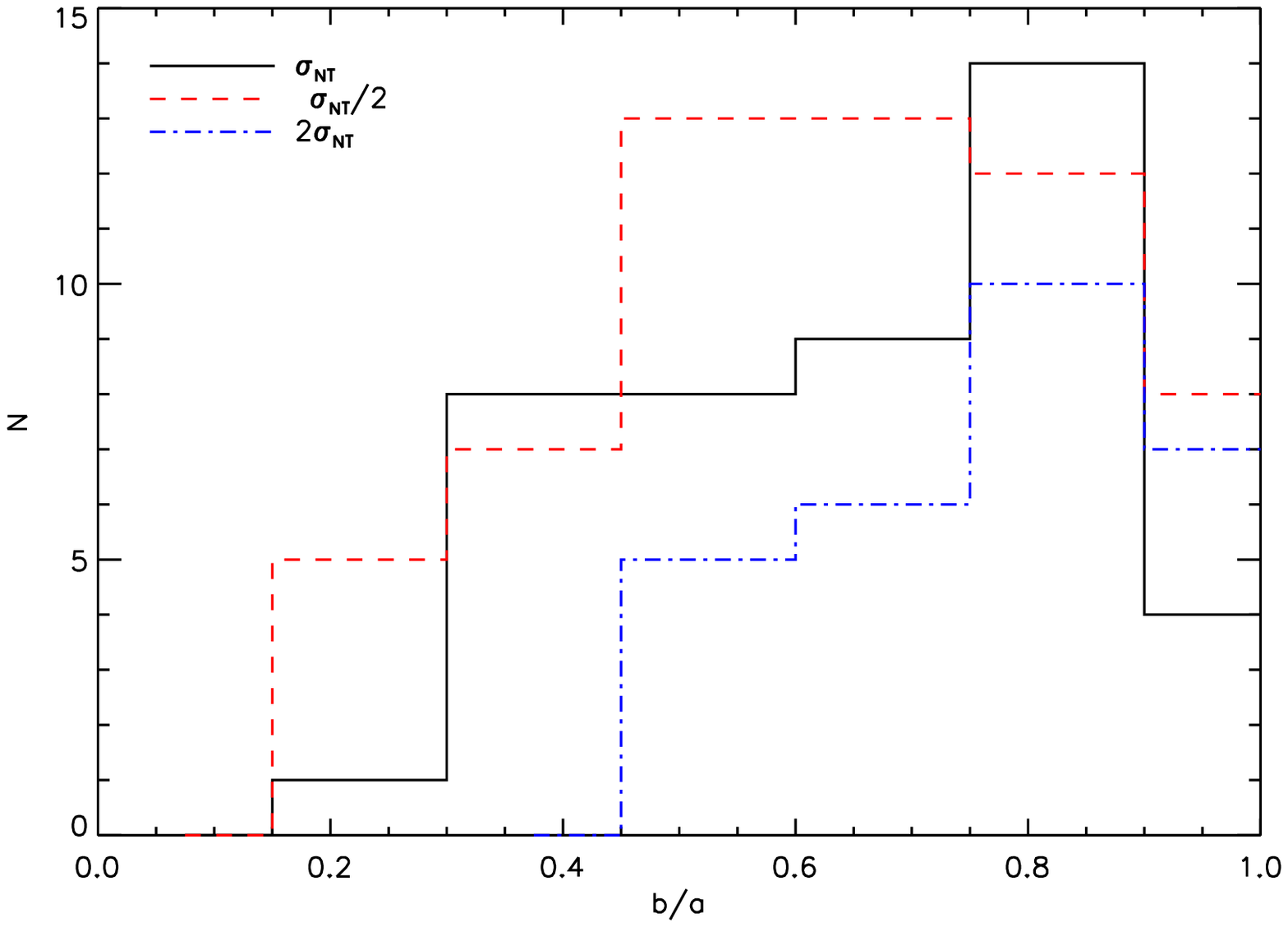} 
\caption{Number of cores as a function of axis ratio $q=b/a$ along a single projection for cores defined using the fiducial $\sigma_{NT}$,  2$\sigma_{\rm NT}$, and 0.5$\sigma_{\rm NT}$ at 1$t_{\rm ff}$. \label{contours}}
\end{figure}

We verify that the distribution of core ellipses is not strongly dependent on the details of the normalization by comparing the  ellipses in the fiducial case with core samples assuming $\sigma = 0.5 \sigma_{\rm NWT}$, and $\sigma = 2\sigma_{\rm NWT}$. 
This is equivalent to adopting the same contour level while changing the average density of the simulation by a factor of 2.  Figure \ref{contours} shows the fiducial distribution of axes ratios and the two distributions with different contours.  
We find that the new axis ratio distributions are consistent with being drawn from the same distribution at 99\% and 17\% confidence, respectively, when comparing samples of equal number. Using a lower or higher contour level respectively increase or decreases, respectively, the number of cores in the population by $\sim$30\%.

\section{ Data Comparison}

In table 1 we report a summary of the statistical properties of our cores, including the size of the sample, the absolute sizes of cores, and their aspect ratios. We compute these properties for all cores, for the real Orion data set, and for both the driven and undriven simulations at our two sample times $t={1\over 2}t_{\rm ff}$ and $t=t_{\rm ff}$. In each case we report properties both for the entire population of cores regardless of whether they contain stars and for the starless and protostellar populations separately.

\subsection{ Core Sizes}

In this section, we compare the physical sizes of the simulated and observed cores, where the lengths of the semi-major and semi-minor axes are given in Table \ref{table1}. 
Generally, we find that the medians of the net distribution of simulated core sizes are $\sim20$\% smaller. Whereas the simulated starless core sizes are fairly similar to observation, the observed protostellar core sizes are much larger than their simulated counterparts. 
This discrepancy is likely an artifact of the sink particle accretion algorithm: once a core forms a sink particle the surrounding gas is accreted more quickly and to higher masses without losses from outflows, so that the reservoir of bound gas around the sink particle is rapidly depleted. 
It is also possible that magnetic fields, which we neglect, play a significant role in supporting the protostellar envelopes, thus slowing the collapse process and contributing to the larger sizes of the observed protostellar cores.

\begin{figure}
\plotone{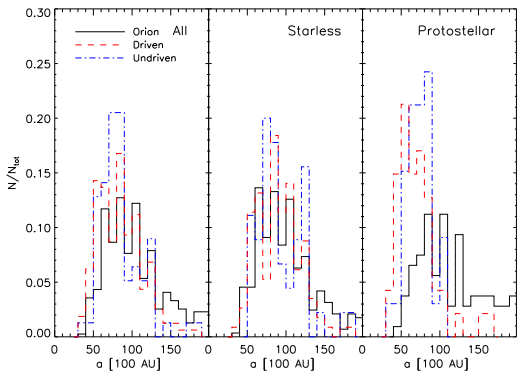} 
\caption{The sizes of major axis, a,  in units of 100 AU for the total, starless, and protostellar cores, from left to right, at 1$t_{\rm ff}$.   \label{size}}
\end{figure}

As shown in figure \ref{size}, the simulation core size distributions have smaller dispersions than the observation. Although some of the difference can be attributed to the smaller sizes of protostellar cores, the smaller core sample and simulation domain size may also contribute. The minimum core size is most likely set by the observation resolution.
%

We can compensate for the rapid depletion of gas in the outer parts of protostellar cores if we adopt a significantly lower estimate of the telescope noise ($\sigma \sim 0.2 \sigma_{\rm NWT}$) and thus a lower contour threshold for defining cores. This has the effect of making the protostellar cores somewhat larger while leaving the starless sizes mostly unchanged, so that the overall size distribution is in better agreement with the observations. Using this lower noise level does not significantly alter the core shape distribution, however, which suggests that the discrepancy in core sizes is not significant for the purpose of determining core shapes. We therefore proceed with our analysis using the real telescope noise level, 
$\sigma = \sigma_{\rm NWT}$.

\begin{deluxetable*}{lccccccccccccccccc} 
\tablewidth{0pt}
\tablecolumns{18}
\tablecaption{Core axis ratio $b/a$ minimum, median, and mean and median core sizes. \label{table1}}
\tablehead{
\colhead{} &
\multicolumn{5}{c}{All} &
\colhead{} &
\multicolumn{5}{c}{Starless} &
\colhead{} &
\multicolumn{5}{c}{Protostellar} \\
\colhead{} & \colhead{O\tablenotemark{a} } &
\colhead{D$_1$\tablenotemark{b} } & \colhead{D$_{{1\over2}}$\tablenotemark{c} }& \colhead{U$_1$\tablenotemark{d} } &\colhead{U$_{1\over2}$\tablenotemark{e} }  &
\colhead{} &
\colhead{O} & \colhead{D$_1$} & \colhead{D$_{{1\over2}}$} &
\colhead{U$_1$} & \colhead{U$_{{1\over2}}$} &
\colhead{} &
\colhead{O} & \colhead{D$_1$} & \colhead{D$_{{1\over2}}$} &
\colhead{U$_1$} & \colhead{U$_{{1\over2}}$} 
}
\startdata
$N_{\rm cores}$ &  393 & 161 & 152  & 78  & 66  & 	
&286 & 114  &  103  & 45  & 50 &    	&107 & 47 &49  & 33 & 16  \\ 

Minimum $b\over a$  &0.24  & 0.22 & 0.23  &0.18 & 0.23 &   	
& 0.24 & 0.22 &0.23 &  0.18&0.23 &	& 0.34&  0.23 & 0.26 & 0.28 & 0.66 \\ 

Median $b\over a$  &0.66  & 0.68 & 0.58  &0.65 & 0.65 & 	
&  0.66& 0.66 & 0.55 & 0.68&0.57&	& 0.68& 0.68 &0.74  &  0.79  & 0.76\\ 

Mean $b \over a$     & 0.67   & 0.66 &  0.61 &0.66 &0.62  &  	
& 0.66& 0.64  & 0.56    & 0.58& 0.57&  	& 0.68 &0.68& 0.73 & 0.77 & 0.80 \\

Median a$_{100}$\tablenotemark{f}&   100  & 76& 84 & 80 & 92&	     &96 &  88 & 88  & 80 & 104&    & 120 &64& 68 & 72& 64 \\  

Median b$_{100}$\tablenotemark{f}&  64   & 48 & 48 & 52 & 56&	   &64 &  54& 48  &48 &   56&    & 76 & 48 & 48 & 52 & 48\\
\enddata
\tablenotetext{a} {Observed Orion molecular cloud cores (NWT)}
\tablenotetext{b} {Driven turbulence simulation at 1$t_{\rm ff}$}
\tablenotetext{c} {Driven turbulence simulation at ${1\over{2}}t_{\rm ff}$}
\tablenotetext{d} {Undriven turbulence simulation at 1$t_{\rm ff}$}
\tablenotetext{e} {Undriven turbulence simulation at ${1\over{2}}t_{\rm ff}$}
\tablenotetext{f} {Median projected semi-major (a) and semi-minor (b) size in units of 100 AU. }
\end{deluxetable*}

\subsection{ Overall Shape Distributions}

As shown in Table \ref{table1}, we find similar means and medians for the shape distributions. The characteristic mean falls around 
$q=0.6-0.7$. 
The maximum aspect ratio is also very similar in all the cases, around 4:1, and the most elongated core is starless. In the simulations, core elongation is a result of the initial filamentary gas structure out of which the cores form. It is the remnant of the turbulence rather than a signature of magnetic fields. 

We next characterize the similarity of the distributions by using a KS test. One interesting aspect of the Orion dataset is that we can rule out the possibility that the separate Orion A and B populations originate from the same parent population with $>95$\% confidence.  This disagreement is caused by the higher fraction of elongated cores in Orion B. The physical difference between the North and South regions are otherwise not large: both are sites of high-mass star formation, both have similar average column densities \citep{mad86}, and the patches surveyed are roughly the same size. However, Orion A has a 25\% larger velocity dispersion, 5.1 km s$^{-1}$ as measured in CO \citep{mad86}.  The magnitude of this difference is a useful number to bear in mind when characterizing the extent of agreement or disagreement with the simulations. In comparison, the driven and decaying samples are consistent with being drawn from the same parent population at $88$\% 
and $66$\% confidence for 1$t_{\rm ff}$ and ${1\over 2} t_{\rm ff}$, respectively.

We note that the KS statistic is somewhat influenced by the size of the distribution being compared. For example the smaller the samples, the more likely the test will conclude two samples are consistent with being drawn from the same distribution. In general we find if we always compare distributions of the same size (by randomly selecting cores from the larger distribution), KS agreement rises by $5-10\%$.

\begin{figure}
\plotone{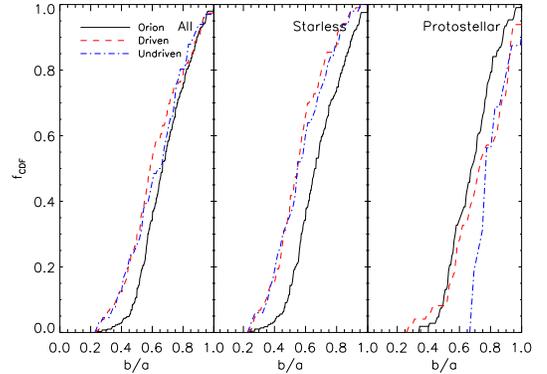} 
\caption{Cumulative distribution function of the total, starless, and protostellar shape distribution, from left to right, at 1$t_{\rm ff}$.   \label{cdf}}
\end{figure}

Table \ref{table2} shows the KS statistics for comparisons of the axis ratios in Orion to each of the simulations. Overall, the decaying turbulence simulation agrees better with the Orion populations, although both simulations have fairly high agreement with Orion B. Given the uncertainties in the normalization and the disagreement between the Orion A and B samples, we consider agreement greater than 10\% to be encouraging. Interestingly, the disagreement of the driven population is determined mainly by the disagreement of ellipses in one particular projection at 1$t_{\rm ff}$, which has an overabundance of elongated cores. Minus the cores in this projection, the driven sample agrees with $\sim17\%$ confidence at this time.  
Figure \ref{cdf} shows the cumulative distribution function of the core shapes for the observations and two simulations.

\begin{deluxetable}{cccc}
\tablewidth{0pt}
\tablecolumns{4}
\tablecaption{KS agreement of simulations with Orion for net populations \label{table2}}
\tablehead{  
\colhead{} & \multicolumn{3}{c}{Orion}  \\
\colhead{} & \colhead{O\tablenotemark{a} total (\%)} & \colhead{A (\%)} & \colhead{B (\%)} 
}
\startdata
{D$_1$\tablenotemark{b}} & 2.3 & 1.2  & 32.9     \\
{D$_{1\over2}$}                    & 0.03 & .004  & 3.7  \\
{U$_1$\tablenotemark{c}} & 25.2 & 18.0 &  67.7 \\
{U$_{1\over2}$} 		& 3.8 & 1.0 &  16.0       \\

\enddata
\tablenotetext{a} {Observed Orion molecular cloud cores (NWT)}
\tablenotetext{b} {Driven turbulence simulation}
\tablenotetext{c} {Undriven turbulence simulation}
\end{deluxetable}

\subsection{ Starless and Protostellar Core Shapes}

KS agreement for starless and protostellar cores is compared in Table \ref{table3}. We find similar agreement with the Orion data for both the driven and undriven 
starless
core shapes at 1$t_{\rm ff}$. In contrast with the observational data, which has a high level of agreement between the Orion starless and Orion protostellar core shapes ($\sim$ 60\%) the driven starless and driven protostellar core shapes are only moderately similar ($\sim$13\%), while the decaying starless and decaying protostellar core shapes are quite dissimilar ($\sim 0.02$\%).  This is illustrated in second and third panels of Figure \ref{cdf}, which show the undriven starless distribution of shapes significantly to the left of the Orion distribution, while the undriven protostellar distribution falls to the right. 
As a result, the individual simulation starless and protostellar shape distributions can be much less similar to the observed cores than the net simulation shape distribution (see $U_1$ in Table \ref{table2} and \ref{table3}).
%
One caveat of this comparison is that the actual goodness of agreement depends not only upon agreement between the net shape distributions but upon the agreement between the individual starless and protostellar shapes and observations.
Figure \ref{cdf} suggests that, particularly for the decaying case, the K-S test may overestimate the overall similarity to observed cores.
Observational data bears out the similarity of the distributions of starless and protostellar shapes \citep{myers91, jijina99}. In the decaying simulation, the difference is likely due to the more rounded protostellar cores, which are experiencing strong collapse \citep{Offner08b}.

\begin{figure}
\plotone{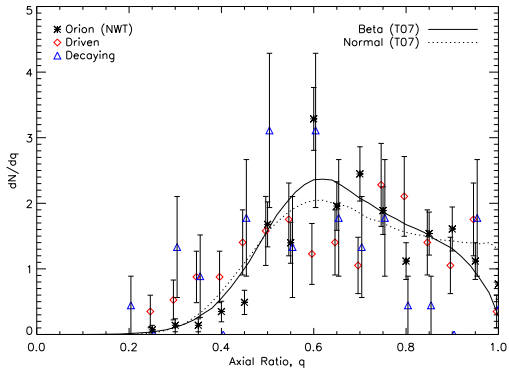} 
\caption{Histrogram of starless cores ratios $q=b/a$ at 1$t_{\rm ff}$. Results from T07 derived assuming underlying beta and normal distribution for core ratios have been overlaid. The error bars shown on each bin reflect $\sqrt{N}$ counting statistics. The bins for each sample are centered at the same values of $q$, but the plotted points have been offset slightly to the left or right to allow the error bars to be distinguished.  \label{dndq}}
\end{figure}

\begin{figure}
\plotone{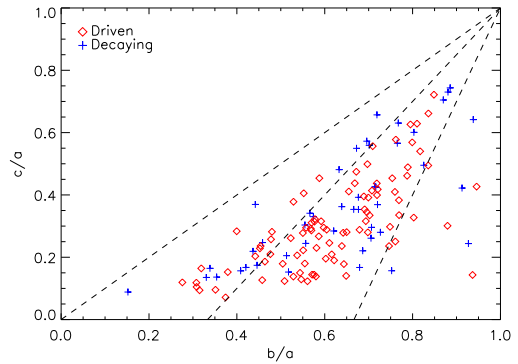} 
\caption{Plot of the 3D core aspect ratios $c/a$ vs. $b/a$ at 1$t_{\rm ff}$. The dashed lines indicate the boundary between prolate, triaxial, and oblate cores, from left to right.  \label{3dshapes}}
\end{figure}

In Figure \ref{dndq} we plot the number of starless cores as a function of axis ratio, $q$. For comparison, we also plot the T07 maximum-likelihood curves drawn from normal and beta distributions.  In spite of the larger variation in the simulation data (due to the smaller sample size), the simulation data also appears to follow the curves reasonably well. 
 \cite{Offner08a} report that the protostellar cores in both simulations are mainly trixial, with some preference for prolateness over oblateness. 
To investigate the 3D shape distribution of the starless cores, which were not examined in \cite{Offner08a}, we first triangulate the 2D projected positions to identify the 3D coordinates of the core center. By setting a minimum density cutoff for cells within 0.1 pc of the core center, we define the gas contained in the core.  We adopt a density cutoff of the minimum of $n_{\rm H} = 2\times 10^4$ cm$^{-3}$ and $0.2~n_{\rm Hpeak}$, where protostellar cores generally satisfy the former and starless cores the latter. 
Unlike the core definition in \cite{Offner08a}, we do not require that the gas be bound. We apply principle component analysis to the set of cells comprising each core to identify the eigenvalues of the cardinal axes \citep{jol02}. As shown in Figure \ref{3dshapes}, the cores for both simulations are predominately triaxial. The remaining cores are preferentially prolate. However, the ratio of the number of prolate to oblate cores in 3D is somewhat sensitive to the chosen cutoff density.
These results appears to be inconsistent with the claim of T07 that the observed core axis distribution implies that prolate cores are rare while oblate cores are more common.

\begin{deluxetable}{cccc}
\tablewidth{0pt}
\tablecolumns{4}
\tablecaption{KS agreement of simulations with Orion for starless and protostellar populations \label{table3}}
\tablehead{
\colhead{} & \colhead{O\tablenotemark{a} Starless (\%)} & \colhead{} & \colhead{O Protostellar (\%)}
}
\startdata
{D$_1$\tablenotemark{b} Starless} & 1.2 & {D$_1$ Proto} & 20.7 \\
{D$_{1\over2}$     Starless    }  & 2.2$\times 10^{-5}$ & {D$_{1\over2}$ Proto}   & 2.5  \\
{U$_1$\tablenotemark{c} Starless} & 1.2 & {U$_1$ Proto} &  2.9 \\	 
{U$_{1\over2}$ Starless} & 0.04 & {U$_{1\over2}$ Proto} &  0.4 \\	 
\enddata
\tablenotetext{a} {Observed Orion molecular cloud cores NWT}
\tablenotetext{b} {Driven turbulence simulation}
\tablenotetext{c} {Undriven turbulence simulation}
\end{deluxetable}

\subsection{ Time Dependence of Core Shapes}

We find less KS agreement between the Orion core shapes and the simulated cores at ${1\over 2} t_{\rm ff}$. The origin of the disagreement is mainly due to the smaller mean axes ratios in both the driven and decaying cases as illustrated in Table \ref{table1}. The difference between starless cores and protostellar cores is also more pronounced at this time since the starless cores in both simulations are more elongated and the agreement between the observed and simulated starless distributions is much  worse than for the protostellar core distributions. This may be because the simulated protostellar cores are necessarily located in regions of the flow that are dominated by gravitational effects. In contrast, at early times the shapes of starless cores are more strongly influenced by turbulence rather than gravity and so they appear more elongated and filamentary.  This suggests that gravitational fragmentation, not only turbulence, significantly influences core shapes.

\section{Conclusions}

Overall, we find a high level of similarity between observations of core axis ratios in Orion and simulations of core formation in a self-gravitating, non-magnetized turbulent medium, with either driven or decaying turbulence with best agreement occurring after one dynamical time. 
The similarity extends from the mean and median core axis ratios to the distributions of starless and protostellar core shapes. We obtain good agreement despite the absence of magnetic fields in the simulations, which may indicate that the local magnetic field in Orion is not strongly influencing the core shapes. 
This is supported by both turbulent magnetic simulations such as those by \citep{ball02} and observations that find the shapes of high density structures are not strongly correlated with the magnetic field direction \citep{ball07}.
Moreover, we find that a population of cores that is intrinsically triaxial, but with a tendency to be more prolate than oblate, can produce an observed distribution of core axis ratios consistent with what is seen in Orion.

The axis ratio distributions are also quite similar in the simulations with driven and decaying turbulence. In fact, we find that the shape distributions of Orion A and Orion B are often more dissimilar (to 95\% confidence) to each other than to the simulations. This indicates that the level of turbulence does not play a significant role in determining core shapes. Both simulations compare with larger confidence to the Orion B core sample, although in terms of statistics, the decaying simulation gets slightly better agreement with the total Orion sample of cores. 

Increasingly large and complete observational data sets invite important comparisons with simulations, which could shed light on both the theory of star formation and details of the molecular clouds we observe. Our results provide a cautionary note that such comparisons should preferably be done by projecting from simulations into the observational domain, including realistic sensitivities and resolutions. Further simulations concerning the effects of magnetic fields, combined with more detailed simulated observations, would be a beneficial future direction of research.

 \acknowledgments{
The authors thank K. Tassis and J. Goldston-Peek for helpful discussions and D. Nutter for details and clarifications concerning his recent work on Orion.
 In addition, helpful suggestions from the referee, J. Ballesteros-Paredes, significantly improved this manuscript.
 Support for this work was provided by the US Department of Energy at the Lawrence Livermore National Laboratory under contracts  B-542762 (S.~S.~R.~O.); NASA through Hubble Fellowship grant HSF-HF-01186 awarded by the Space Telescope Science Institute, which is operated by the Association of Universities for Research in Astronomy, Inc., for NASA, under contract NAS 05-26555 (M.~R.~K.).
 Computational resources were provided by the NSF San Diego Supercomputing Center through NPACI program grant UCB267; and the National Energy Research Scientific Computer Center, which is supported by the Office of Science of the U.S. Department of Energy under contract number DE-AC03-76SF00098, though ERCAP grant 80325.}


\end{document}